\documentclass[final,3p,times,twocolumn]{elsarticle}

\usepackage{graphics}
\usepackage{graphicx}
\usepackage{epsfig}
\usepackage{epstopdf}
\usepackage{amssymb}
\usepackage{amsthm}
\usepackage{amsfonts}
\usepackage{amsmath}
\usepackage[colorlinks={true}]{hyperref}
\usepackage{balance}

\newcommand{\ket}[1]{|#1\rangle}
\newcommand{\bra}[1]{\langle #1|}

\begin{document}

\begin{frontmatter}



\title{A steady state quantum classifier}
\ead{dturkpence@itu.edu.tr}
\author[1]{Deniz T\"{u}rkpen\c{c}e}
\author[1]{Tahir \c{C}etin Ak{\i}nc{\i}}
\author[1]{Serhat \c{S}eker}
\address[1]{Department of Electrical Engineering, \.{I}stanbul Technical University, 34469 \.{I}stanbul, Turkey}


\begin{abstract}

We report that under some specific conditions a single qubit model weakly interacting with information environments can be referred to as a quantum classifier. We exploit the additivity and the divisibility properties of the completely positive (CP) quantum dynamical maps in order to obtain an open quantum classifier. The steady state response of the system with respect to the input parameters was numerically investigated and it's found that the response of the open quantum dynamics at steady state acts non-linearly with respect to the input data parameters. We also demonstrate the linear separation of the quantum data instances that reflects the success of the  functionality of the proposed model 
both for ideal and experimental conditions. Superconducting circuits were pointed out as the physical model to implement the theoretical model with possible imperfections.

\end{abstract}


\end{frontmatter}



\section{Introduction}

Classification of data is of central importance to important implementations such as medical diagnosis, pattern recognition and machine learning. Due to the well-known advantages of quantum computation, studies about the quantum equivalent of machine learning algorithms have been reached to a remarkable level \cite{banchi_quantum_2016, rebentrost_quantum_2014, schuld_quantum_2018, schuld_simulating_2015,yamamoto_simulation_2018, lu_separability-entanglement_2018, lloyd_quantum_2018}. In contrast to the circuit model of quantum computation in which the system of interest is assumed to be perfectly isolated from the environmental degrees of freedom, one could imagine a quantum classifier as an open quantum system. This model could be referred to as a quantum data driven decision making process, as the environmental states carry information content. 

Recent studies underline that the quantum reservoirs are not necessarily the rubbish bins in which the useful information is lost, but they are communication channels that they transmit information~\cite{zwolak_redundancy_2017, blume-kohout_simple_2005}. Moreover, the proposed quantum equivalent of learning schemes are reported to dynamically violate the unitarity even for a minimal classifier level~\cite{schuld_simulating_2015}. These facts motivate us to study the possibility of the basic quantum classifiers in the context of open quantum systems in which the dynamics are non-unitary. 

 In this study, we numerically demonstrate that a steady state of a quantum unit subjected to different information environments acts as a quantum data classifier. We investigate a model that represents mixing properties of quantum dynamical maps and demonstrate that the mixture of quantum dynamical maps can be natural data classifiers under some circumstances. The influence of a dissipative environment on the reduced system dynamics is that the evolution of pure states into mixed steady states~\cite{breuer_theory_2007}.  Mixed quantum states are mixtures of classical probability distributions carry no quantum signature. Therefore quantum mixed states seem useless for quantum computing implementations. However, it's possible to demonstare most of the quantum apllications by mixed state~\cite{siomau_quantum_2011} or dissipative quantum computing~\cite{verstraete_quantum_2009}. 
 
\begin{figure*}[!t]
\includegraphics[width=6.4in]{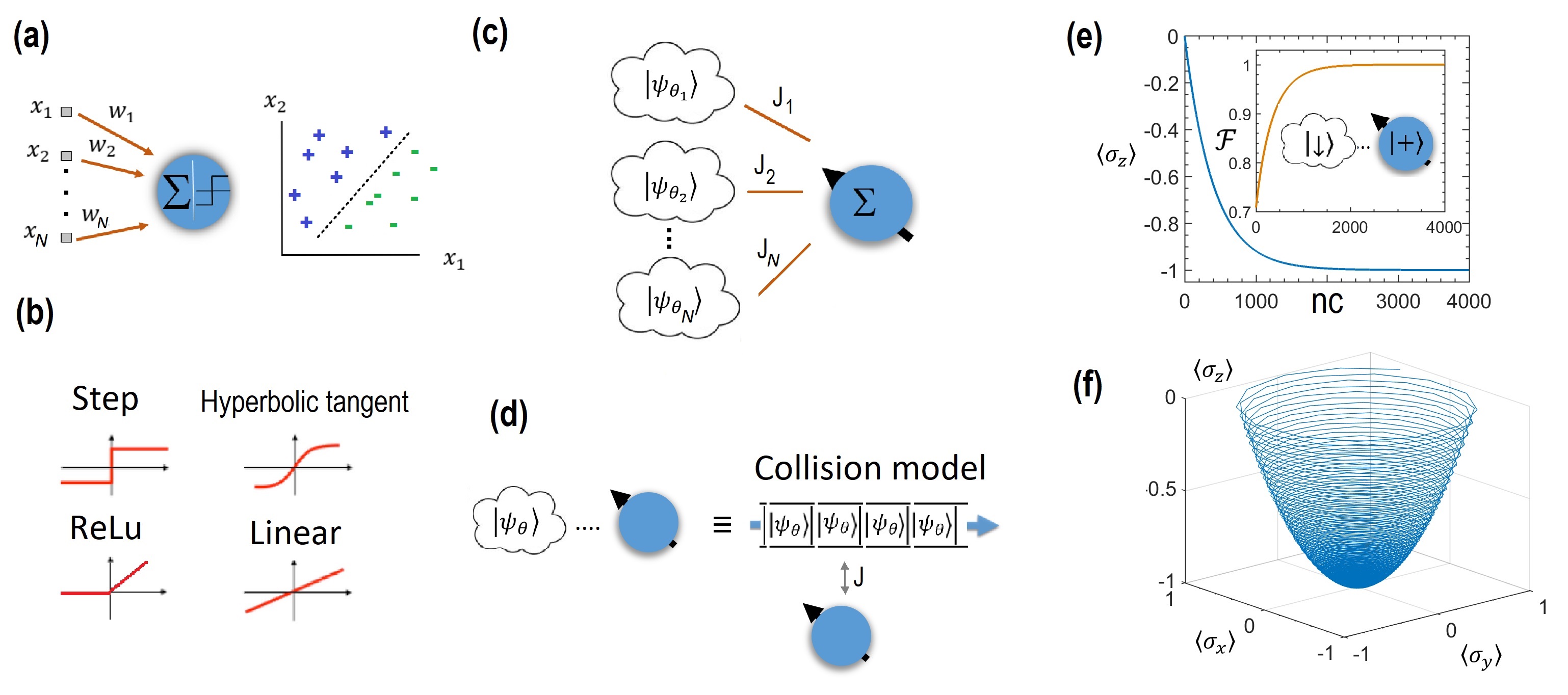}
\caption{ \label{fig:Fig1} (Colour online.) A general view of the proposed methods. (a) A classical perceptron with $N$ inputs. (b) A few of the activation functions for perceptrons. (c) The scheme of the proposed quantum classifier. A single spin is weakly coupled to n number of reservoirs carrying quantum information. (d) The collision model to simulate the open quantum system dynamics. (e) Time evolution of the single spin magnetization depending on the number of collisions (nc). The single spin was initially prepared in $\ket{+}$ state and contacted with a single  reservoir spin-down reservoir. Also the time dependent fidelity (inset) between the qubit and the fixed reservoir state has been depicted. (f) The Bloch ball vector trajectory of the single spin during the evolution. The coupling between the each environment unit and the single spin is $J=0.1$. The duration of the each unit interaction between the units and the spin is $\tau = 5\times 10^{-2}/J $.} 
\end{figure*} 

Exploiting the quantum resources exhibit quadratic or non-linear response against linear variation of resource parameters is of importance to quantum thermodynamic or material sciences \cite{turkpence_quantum_2016, turkpence_photonic_2017, turkpence_engineering_2014}. In our model, we focus on a single spin weakly coupled to information reservoirs and calculate the reduced dynamics by tracing out the environmental degrees of freedom in Markov approximation. Single spin magnetization is the figure of merit as the steady state response of the reduced dynamics. In the model, information reservoirs connected to the single spin represents the input data. We find that the steady state response of the model varies non-linearly with respect to the linear variation of input data just like the activation functions of the classical classifiers. The underlying physics of our model relies on the complete positivity, additivity~\cite{kolodynski_adding_2018} and divisibility~\cite{wolf_dividing_2008, filippov_divisibility_2017} of quantum dynamical maps. We limit our scope with the classification analysis and leave how to perform the training and learning of the model, to an else study. 

\section{Framework and system dynamics}\label{sec:Sc1}

The simplest mathematical model for data classification is a perceptron (see Fig.~\ref{fig:Fig1} (a) left) that predicts an output for a weighted summation of an input data set depending on an activation function. The input dataset (the features)  $x_1, x_2\ldots x_N$ are any measurable individuals with their corresponding adjustable weights $w_1,w_2,\ldots w_N$ and the linear summation $y=\sum_i x_iw_i$ is inserted into an activation function $f(y)$ that returns an output prediction ~\cite{hecht-nielsen_neurocomputing_1990}. Figure~\ref{fig:Fig1} (b) depicts a few of commonly used activations functions. For instance, a step function yields $f(y)=1$ if $y=\sum_i x_iw_i\geq 0$ and yields $f(y)=-1$ else. After these results, if a line correctly separates the data instances (as in Fig.~\ref{fig:Fig1} (a) right), this corresponds to a properly functioning perceptron. One can choose activation functions either with linear or non-linear responses but non-linear functions are appealing for multi-layer neural network applications. 

There are various reports for quantum models of perceptrons or neural networks \cite{banchi_quantum_2016, schuld_quantum_2018, schuld_simulating_2015, yamamoto_simulation_2018} relying on the advantages of quantum computing. Generally these schemes require computational resources proportional with the number of input instances to mimic the activation functions \cite{schuld_simulating_2015}. Our scheme is an open quantum system and we transform the dissipative processes into an advantage for data classification. The input data are the quantum information units characterized by qubits which are refered to as information reservoirs~\cite{deffner_information_2013, mandal_maxwells_2013, strasberg_quantum_2017}. A qubit is parametrized by polar and azimuthal angles as $\ket{\psi\left(\theta,\phi\right)}=\cos\left(\frac{\theta}{2}\right)\ket{\uparrow}+e^{i\phi}\sin\left(\frac{\theta}{2}\right)\ket{\downarrow}$ in the well-known Bloch sphere representation. Throughout of our study we take $\phi=0$ fixed and parametrize `quantum features' by $\theta$. In the calculations, we use radians and degrees interchangeably. We present our classifier as a model in which a single spin is weakly coupled to different reservoirs carrying information content. We adopt a repeated interaction process to model the open quantum dynamics~\cite{bruneau_repeated_2014}. Repeated interactions that are also known as collision models have became very popular recently due to their flexibility to choose the associated reservoir states and find applications to model non-Markovian~\cite{cakmak_non-markovianity_2017} as well as Markovian and quantum correlated reservoirs~\cite{lorenzo_composite_2017}.
 
As depicted in Fig.~\ref{fig:Fig1} (d), initially prepared identical ancillas $\lbrace \mathcal{R}_n \rbrace$ sequentially collides with the system $\mathcal{S}$ with equal duration $\tau$. It's assumed that initially, system plus reservoir $\mathcal{SR}$ state is in a product state $\varrho(0)=\varrho_{\mathcal{S}}(0)\otimes \varrho_{\mathcal{R}}$ where $\varrho_{\mathcal{S}}(0)=\ket{+}\bra{+}$ and $\varrho_{\mathcal{R}}=\ket{\psi_{\theta}}\bra{\psi_{\theta}}$. We choose the initial system states as $\ket{+}=(\ket{\uparrow}+\ket{\downarrow})/\sqrt{2}$ in order to  provide a null magnetization initially. In this study, we use standard collision model in which the ancillas do not interact each other, hence the open system evolution is Markovian. The collisions between the system qubit and the each ancilla are described by unitary propagators $\mathcal{U}_{\mathcal{SR}_n}=e^{-i\mathcal{H}_{\mathcal{SR}_n}\tau}$ where the reduced Planck constant was set $\hslash=1$ throughout the manuscript. $\mathcal{H}_{\mathcal{SR}}$ is the time-independent reservoir ancilla plus system Hamiltonian where 
\begin{equation}\label{Flip}
\mathcal{H}_{\mathcal{SR}_n}=\frac{h}{2}(\sigma_z^n+\sigma_z^s)+J(\sigma_{+}^n\sigma_{-}^s+h.c.).
\end{equation}
Here, $\sigma_z^n$ and $\sigma_{\pm}^n$ are the Pauli matrices acting on the $n^{th}$ ancilla of the reservoir, $\sigma_{\pm}^s$ are the Pauli matrices acting on the system qubit, $J$ is the coupling between the system and the $n^{th}$ ancilla and $h$ is the characteristic frequencies of the system and each ancilla. 

The defined interactions above, give rise to a dynamical map such that 
\begin{equation}
\Phi_{\mathcal{SR}}[\varrho]=\mathcal{U}_{\mathcal{SR}}\left(\varrho_{\mathcal{SR}}^0\right)\mathcal{U}_{\mathcal{SR}}^{\dagger}
\end{equation}
where $\mathcal{U}_{\mathcal{SR}}$ is composed of cascaded applications of $\mathcal{SR}$. In our Markovian scheme, the system of interest evolves into a state identical to the state of the ancillas after sufficient number of collisions. This discrete dynamical process is called quantum homogenization~\cite{scarani_thermalizing_2002}, that is, the system reaches a steady state as 

\begin{align}
\varrho_{\mathcal{S}}^n=&\text{Tr}_n   \big[ \mathcal{U}_{\mathcal{SR}_n}\ldots\text{Tr}_1[\mathcal{U}_{\mathcal{SR}_1}\left(\varrho_{\mathcal{S}}^0\otimes\varrho_{\mathcal{R}_1}\right)\mathcal{U}_{\mathcal{SR}_1}^{\dagger}]\otimes\ldots \nonumber \\ 
&\ldots\otimes\varrho_{\mathcal{R}_n}\mathcal{U}_{\mathcal{SR}_n}^{\dagger} \big]
\end{align}
for sufficiently large number of collisions $n$ where $\text{Tr}_i$ is the partial trace over $i^{th}$ ancilla. The above cascaded dynamical maps can also be presented as 
\begin{equation}
\varrho_{\mathcal{S}}^n=\mathcal{E}_n\circ\mathcal{E}_{n-1}\circ\ldots\circ\mathcal{E}_1\equiv\mathcal{E}^n[\varrho_{\mathcal{S}}^0]
\end{equation}
where $\mathcal{E}^i[\varrho_{\mathcal{S}}]=\text{Tr}_i[\mathcal{U}_{\mathcal{SR}_i}\left(\varrho_{\mathcal{S}}\otimes\varrho_{\mathcal{R}_i}\right)\mathcal{U}_{\mathcal{SR}_i}^{\dagger}]$. Note that each map preserves the density matrix properties such as trace unity and complete positivity, that is, each dynamical map written sequentially above is a completely positive trace preserving (CPTP) dynamical map. 
If a map satisfying $\Phi_{t+s}=\Phi_t\circ\Phi_s$ is CP for all $t$ and $s\geq 0$ then it is a CP divisible map~\cite{wolf_dividing_2008}. Therefore, in this manuscript the standard collision model in which the ancillas are identical and independent, clearly corresponds to CP divisible maps.  
Moreover it's been reported that a collision model can effectively simulate a Markov master equation 
$\partial_t\varrho=\mathcal{L}_t[\varrho]$ as long as it holds the condition of CP divisibility~\cite{filippov_divisibility_2017}. 

As a benchmark calculation, we contact the single spin to a data reservoir in the   $\rho_{\pi}=\ket{\downarrow}\bra{\downarrow}$ fixed quantum state and apply the above formulation as in Fig.~1 (d). We observe that the time time evolution of spin magnetization converges to $\langle\sigma_z(t)\rangle=-1$ as the spin density matrix approaches to the unit fidelity $\mathcal{F}(t)=\text{Tr}\sqrt{\sqrt{\rho_{\pi}}\varrho_{\mathcal{S}}(t)\sqrt{\rho_{\pi}}}=1$ with the fixed reservoir state monotonically. Fig. 1 (e) illustrates the Bloch vector trajectory during the evolution in terms of the statistics of typical observables. By these numerical results, one concludes that our standard repeated quantum interaction process (collision model) can faithfully simulate the CP divisibility and open quantum dynamics in Markov approximation.

\section{Results}
\subsection{Theoretical model}
 In this subsection we present the theoretical modelling of the proposed classifier without accounting for the imperfections or the physical decay mechanisms. The objective is to demonstrate that a small quantum system weakly in contact with different quantum environments can be used for classifying the data in which the environments contain. By `small', it's implied that the system is small enough to be equilibrated toward a steady value in the long term limit~\cite{breuer_theory_2007, linden_quantum_2009}. To this end, the system of interest is weakly coupled to multiple reservoirs as in Fig.~\ref{fig:Fig1} (c). In this scheme, the dynamical evolution can be presented as the mixture of CP divisible dynamical maps
\begin{equation}~\label{Eq:convex}
\Phi_n=q_1\Phi^1_n+q_2\Phi^2_n+\ldots + q_N\Phi^N_n
\end{equation}
by considering their linear convex combinations. Here, $q_i\geq 0$ and $\sum_i^N q_i=1$. Eq.~(\ref{Eq:convex}) is the mathematical description of the implementation of the proposed open quantum classifier in contact with $N$ reservoirs. It's known that Eq.~(\ref{Eq:convex}) can also be presented by a master equation composed of weighted combination of effective generators 
\begin{equation}\label{Eq:ConvexL}
\frac{\partial \varrho}{\partial t}=P_1\mathcal{L}^{(1)}_t+\ldots +P_N\mathcal{L}^{(N)}_t
\end{equation}
again depending on the condition that each generator holds the CP divisibility~\cite{filippov_divisibility_2017} and weak coupling to the reservoirs~\cite{kolodynski_adding_2018}. 
Here, $P_i$ are the probabilities of the system experiencing from the $i^{th}$ environment. 

\begin{figure}[!t]
\includegraphics[width=3.1 in]{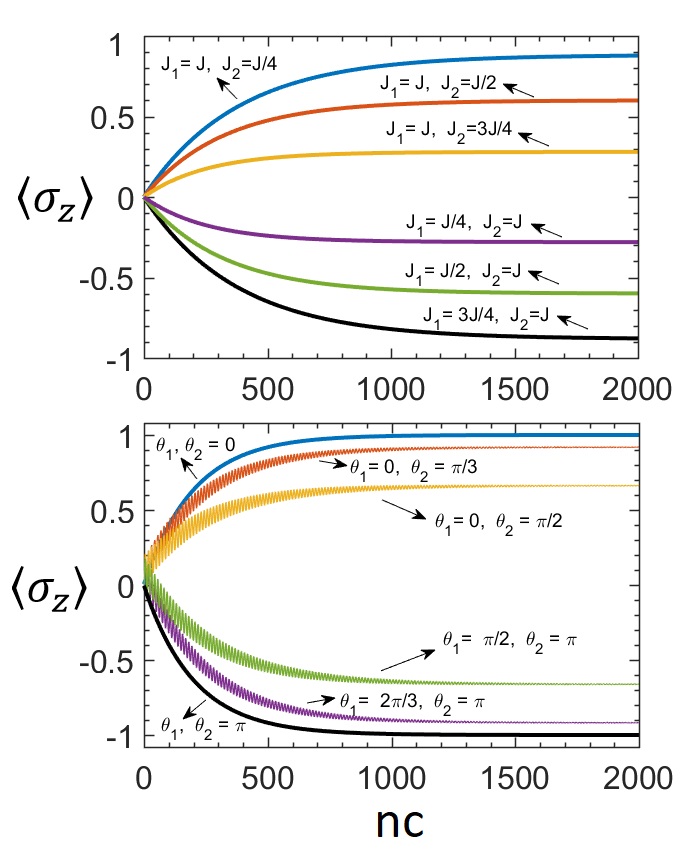}
\caption{ \label{Fig:fig2} (Colour online.) Mixing quantum dynamical maps and evolution of single spin towards steady state by two reservoirs depending on the number of collisions (nc). (a) The state of two reservoirs are fixed and $\ket{\uparrow}$ and $\ket{\downarrow}$ respectively. The evolution of spin magnetization depending on different couplings to reservoirs depicted. (b) The coupling of the spin to the reservoirs are fixed, equal and $J=0.1$. The evolution of spin magnetization depending on different reservoir (qubit) states are depicted. The duration of each interaction between the ancillas and the system qubit is $ \tau = 5\times 10^{-2}/J$. } 
\end{figure}

As mentioned above, the steady state magnetization $\langle \sigma_z\rangle_{ss}=\text{Tr}[\sigma_z\varrho_{ss}]$ is evaluated as the steady state response of the system  for the classification process. Since the system of interest is only a single qubit, the steady state can be defined as a mixed state  $\varrho_{ss}=\sum_i p_i\Pi_{\theta_i}$  where $\Pi_{\theta_i}=\ket{\theta_i}\bra{\theta_i}$ are rank-one projectors stands for orthogonal basis states $\ket{\theta_i}$ with $\theta_i=0,\pi$.  Here, the corresponding steady state probabilities of the two-level system can be simply referred to as $p_e$ and $p_g$ for $\theta_i=0$ and $\theta_i=\pi$ respectively. At this final state the classification emerges with $class_1$ if $\langle \sigma_z\rangle_{ss}=p_e-p_g\geq 0$ and with $class_2$ else, depending on the states of the quantum reservoirs and the weighted couplings of the system to the reservoirs. 

Before demonstrating the classification process, we show some results presenting the steady state dynamics. In contrast to Fig.~\ref{fig:Fig1} (c) for simplicity, we choose only two information reservoirs with states $\ket{\psi_{\theta_1}}$ and $\ket{\psi_{\theta_2}}$ connected to our single qubit system by dipolar $J_1$ and $J_2$ couplings. We consider two cases in our calculations; first the reservoir states are fixed (and orthogonal) and the couplings are varied. Second, the couplings are fixed (and equal) and the reservoir states are varied. Fig. 2 presents the results for these two cases. In Fig.~\ref{Fig:fig2} (a) the single qubit with initial $\ket{+}$ state is connected to the two reservoirs with fixed $\ket{\psi_{\theta=0}}\equiv \ket{\uparrow}$ and $\ket{\psi_{\theta=\pi}}\equiv \ket{\downarrow}$ states with corresponding $J_1$ and $J_2$ couplings respectively. In this case, the evolution of the qubit magnetization toward steady state is depicted with respect to the variation of the $J$ couplings. In the latter case, the variation of the two reservoir states are parametrized by $\theta$ qubit azimuthal angle. As in Fig.~\ref{Fig:fig2} (b) at the initial steps of the evolution for $\theta_i\neq 0$ or $\theta_i\neq \pi$, highly oscillatory behaviour is evident due to the corresponding non-equilibrium reservoir states. However, in both Fig.~\ref{Fig:fig2} (a) and (b) a steady state spin magnetization has been observed after sufficient number of collisions. Note that the couplings are weak and all the conditions for CP divisibility are fulfilled during the evolutions. 

After we confirm that our scheme representing the open quantum dynamics is capable of obtaining the steady states, next we examine the steady state response of the system under linear variation of the input parameters. Again, we consider the two distinct cases; one with the fixed reservoir parameters and the other one with the fixed coupling parameters and again choose spin magnetization as a steady state identifier. 
\begin{figure*}[!t]
\includegraphics[width=6.7 in]{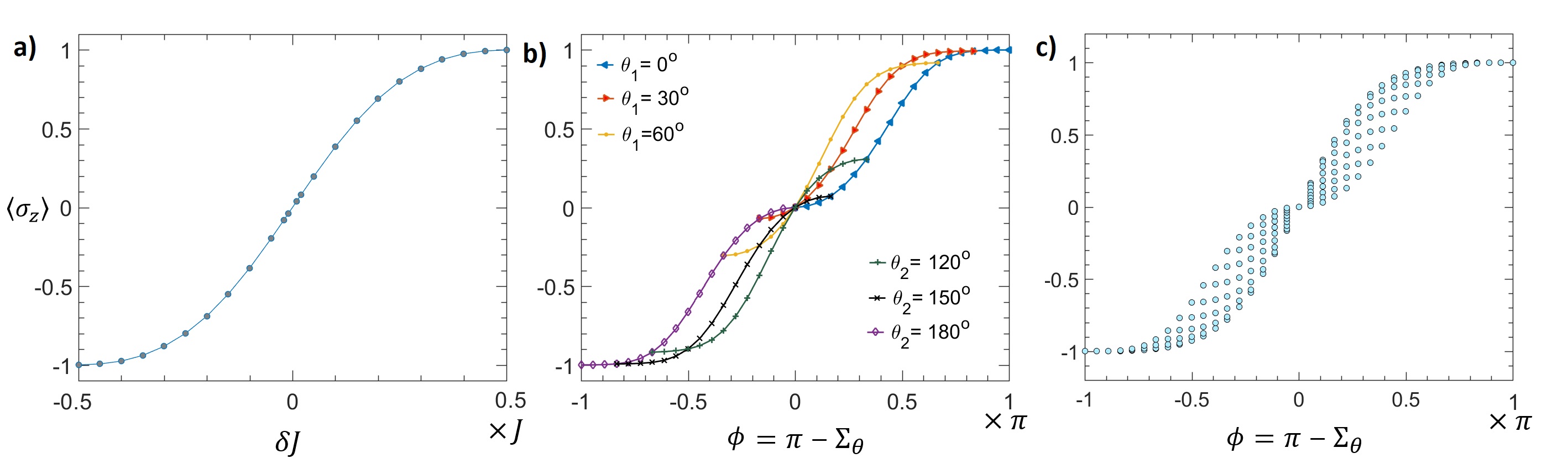}
\caption{ \label{fig:Fig3} (Colour online.) The steady state response of the system depending on the linear variation of the input parameters. (a) The variation of the steady state magnetization of the system qubit depending on the $J_1=J/2+\delta J$ and $J_2=J/2-\delta J$ coupling coefficients where $\delta J$ is a fraction of $J$ with $J=0.1$. The state of the two reservoirs are fixed and $\ket{\uparrow}$ and $\ket{\downarrow}$ respectively. (b) The variation of the steady state magnetization of the system qubit coupled to the two environments carrying different information contents parametrized by $\theta$. In the figure, six curves are plotted with each point representing the steady state magnetization during the presence of two $\ket{\theta_1}$ and $\ket{\theta_2}$ environmental states. Three of the curves (up-right) stand for the three different fixed states of the first environment represented by three azimuthal angles $\theta_1=30^{o},60^{o},90^{o}$ (in degrees).
Each of these curves are composed of 19 points representing the variation of the state of the second environment parametrized by $\theta_2=0^{o},10^{o}\ldots 180^{o}$. Likewise, the remaining three of the curves (down-left) stand for the three  fixed states of the second environment represented by $\theta_2=120^{o},150^{o},180^{o}$ which are composed of 19 points representing the variation of the state of the first environment parametrized by $\theta_1=0^{o},10^{o}\ldots 180^{o}$. Coupling of the system to the reservoirs are fixed, equal and $J_1=J_2=0.1$. 
(c) The variation of the steady state magnetization of the system qubit coupled to the two environments carrying different information contents parametrized by $\theta$. There are $19\times 19 = 361$ plotted dots with each point representing the steady state magnetization during the presence of the first and the second environments, each represented the $\theta=0^{o},10^{o}\ldots 180^{o}$ azimuthal angles. Coupling of the system to the reservoirs are fixed, equal and $J_1=J_2=0.1$.  The magnetization plotted against $\phi=\pi-(\theta_1+\theta_2)$ in both (b) and (c) for convenient scaling as explained in the text.} 
\end{figure*}

Fig.~\ref{fig:Fig3} (a) presents the first case in which the steady state magnetization depicted against $\delta J$ which is a factor governs the variation of the couplings such as $J_1=J/2+\delta J$, $J_2=J/2-\delta J$  to the $\ket{\uparrow}$ and $\ket{\downarrow}$ reservoirs respectively. For instance, when $\delta J=J/2$; $J_1=J$ and $J_2=0$, that is, $\langle \sigma_z\rangle_{ss}=+1$ since the system is coupled only to the first reservoir. As obvious in the figure, the the steady state response of the system is not linear against the overall variation of $\delta J$ and exhibits an activation function-like behaviour such as one of the plots of Fig.~\ref{fig:Fig1} (b). 

In the latter case, the couplings are equal and fixed and the steady response of the system is investigated for different reservoir states defined by the geometrical qubit parameters. In this case, the preferred parameter is the Bloch ball azimuthal angle $\theta$ to define to information reservoir states $\ket{\psi_{\theta}}$. Fig.~\ref{fig:Fig3} (b) depicts the steady response of the system with respect to the variation of one of the two reservoir states while the other one is fixed. Here, the steady response was plotted against $\phi=\pi-\sum_{\theta}$ where $\sum_{\theta}$ is the sum of the two qubit angles $\theta$ representing the reservoir states. For instance, according to the calculations, the steady state of our single qubit system with equal  dipolar couplings to the two reservoirs with orthogonal  $\ket{\psi_{\theta=0}}$ and $\ket{\psi_{\theta=\pi}}$ states is a maximally mixed qubit state with zero magnetization. Hence, $\langle\sigma_z\rangle_{ss}=0$ corresponds to $\phi=\pi-(0+\pi)=0$. Likewise, one obtains the same conditions and maximally mixed qubit state in Fig.~\ref{fig:Fig3} (a) when $\delta J=0$. Fig.~\ref{fig:Fig3} (c) is another way of presenting Fig.~\ref{fig:Fig3} (b) with more steady state points as explained in the caption. Non-linear response of open quantum systems was reported for finite temperature quantum reservoirs~\cite{romero_is_2004}, however temperature is not relevant to our study concerning the information reservoirs. 

\begin{figure*}[!htb]
\includegraphics[width=6.4 in]{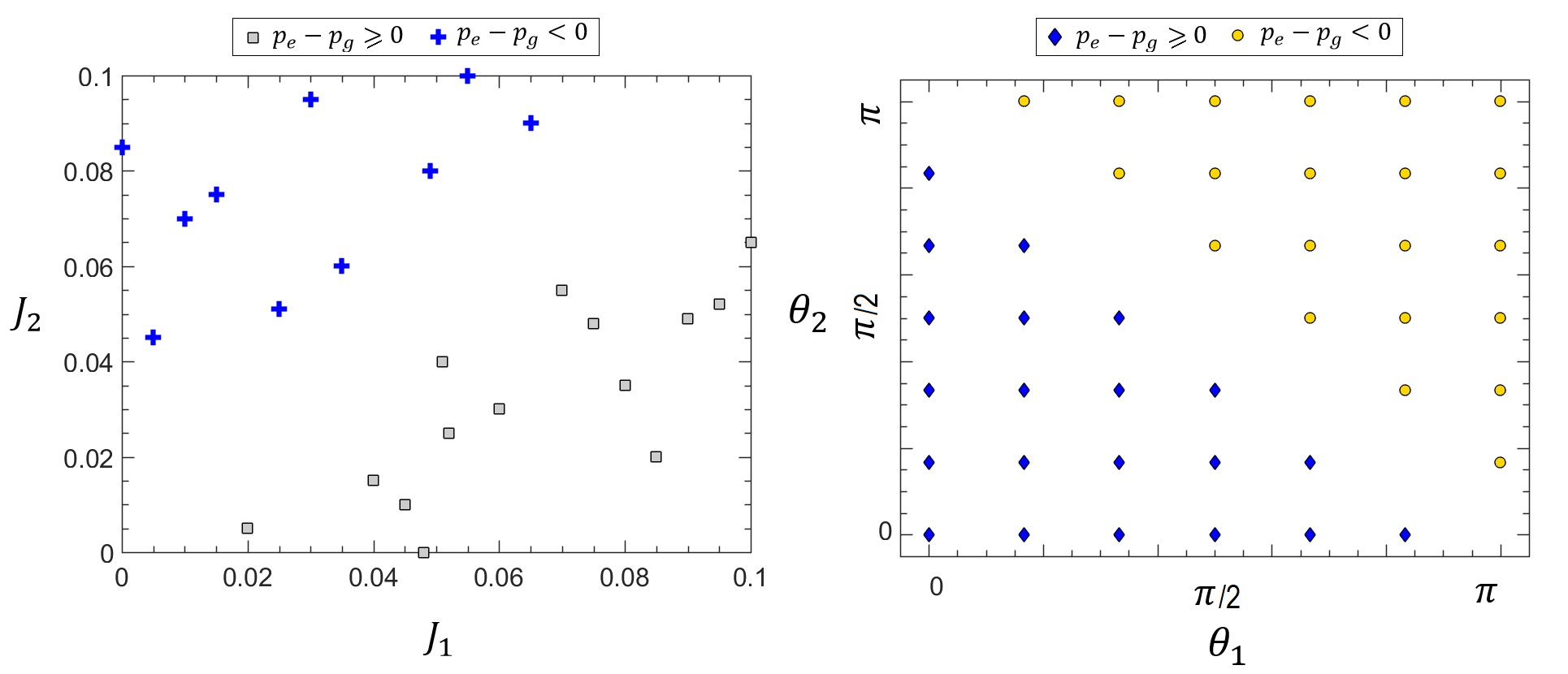}
\caption{ \label{fig:Fig4} (Colour online) Classification by the response of the steady state magnetization of the system qubit for two dimensional inputs.  (a) Classification for the case 1 set up in which the reservoir states are fixed. There are 24 coupling coefficient pairs and the classified instances are linearly separable. (b) Classification for the case 2 set up in which the couplings to the reservoirs are fixed. There are 42 pairs of $\theta$ angles denoting the reservoir states and the classified instances are linearly separable.} 
\end{figure*}

Finally, we illustrate the classification of the input parameters as an examination of the functionality of the quantum classifier. Figs.~\ref{fig:Fig4} (a) and (b) shows that the proposed quantum classifier in the present manuscript is able to linearly separate the input instances composed of the parameters denoting the reservoir states and the couplings to the reservoirs. Bloch sphere representation is very illustrative to represent any two level quantum system. In principle, any point on the sphere is a valid quantum data represented by a pure state. It's well-known that geometrical Bloch sphere representation is parametrized by azimuthal and polar angles $\theta$  and $\phi$. The $\theta$ parameter governs the variation of the qubit state in terms of two orthogonal states while $\phi$ denotes the variation of coherent superposition states. As the classification decision is encoded in the steady state of the classifier in which contains no coherence, we chose to parametrize the initial quantum data of the input channels by $\theta$. As the classical learning algorithms are based on the modification of the weights of the data, visualizing the classification of data in the weight space is quite frequent in classical neuro-computing. Likewise, coupling coefficient ($J$) of the system qubit to the relevant reservoir is the quantum analogue of the classical weights. Therefore we also choose to present the classification plots in the $J$ space. 

By this simple demonstration, it's shown that open dynamics of a single qubit is capable of processing input data in the steady state limit. Though the demonstration is limited to two inputs, extension to arbitrary number of inputs is straightforward due to the convexity of the dynamical maps. As another interesting result, the steady state non-linear response of the system against the linear variation of input parameters encourages one to expand the study toward multi-layer extension of the proposed quantum classifier. 

\subsection{Three input channels}

\begin{figure}
\includegraphics[width=3.1 in]{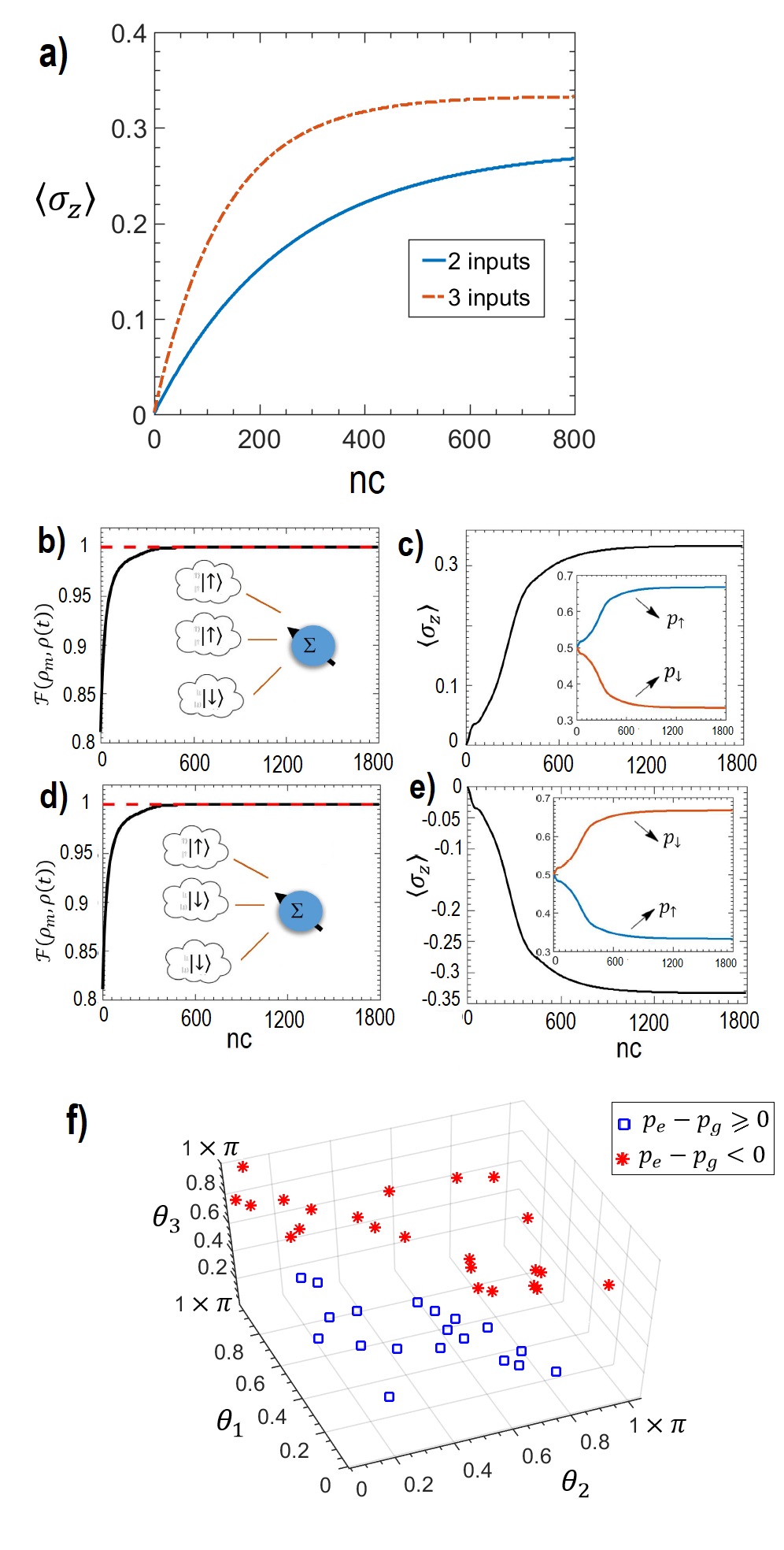}
\caption{ \label{fig:Fig5} (Colour online) Analysis of the classifier with two and three input information channels. Evolution of the spin magnetization of the system qubit for (a) two and three reservoirs and only for 3 reservoirs (c),(e) for different input states depending on the number of collisions presented. In (a), the evolution presented with 2 and 3 input reservoirs with $\ket{\uparrow}$ and $\ket{\downarrow}$ states. For two input case, the states are $\ket{\uparrow}$ and $\ket{\downarrow}$ with corresponding couplings, respectively, $J=0.1$ and $J=0.075$ and for three input case the reservoir states are $\ket{\uparrow}$, $\ket{\uparrow}$ and $\ket{\downarrow}$ with corresponding equal couplings $J=0.1$. In (b)-(f) three reservoirs were considered with (b), (c); $\ket{\uparrow}$, $\ket{\uparrow}$ , $\ket{\downarrow}$ states and (d), (e); $\ket{\uparrow}$, $\ket{\downarrow}$, $\ket{\downarrow}$ states. The time dependent fidelities of the system qubit was calculated (b), (d) where the target state $\rho_m$ denotes the mixture or the linear combination of the reservoir states. The system qubit magnetization and the corresponding diagonal elements were also plotted (c,e) during the evolution. (f) Three dimensional visualization of the classified instances as the steady state response of the system qubit corresponding to three reservoir states denoted by $\theta$. There are 42 triples of $\theta$ angles generated randomly between 0 and $\pi$ with mean 0 and variance 1. (b)-(f) Coupling of the system to the reservoirs are fixed, equal and $J_1=J_2=J_3=0.1$ } 
\end{figure}

 The proposal in which we stress that a single qubit is a binary classifier in the steady state limit was demonstrated for two information reservoirs in the preceding subsection. In the mathematical model, the qubit always returns a binary decision regardless of the number of reservoirs acting as the input information channels as implied in Eqs. (\ref{Eq:convex}) and (\ref{Eq:ConvexL}). Though generalizing the proposal to larger number of input channels are straightforward due to the additivity of the quantum dynamical maps and the convexity of the density matrix, nevertheless, we give an example for three reservoir states as input information channels for further analysis as depicted in Fig.~\ref{fig:Fig5}. 

Before analysing the three channel character alone, we compare the dynamics with the two channel one. A speed up is obvious on the equilibration dynamics of the classifier for three input channels comparing with the two channel input case. One can observe this qualitatively in Fig. ~\ref{fig:Fig5} (a) as the spin magnetization curve saturation for three input case takes place before the two input case. Recent reports support this result by analytical expressions \cite{turkpence_tailoring_2017, manatuly_collectively-enhanced_2018}. That is, unlike the intuitive expectations, the classifier is faster as the new input channels are introduced. This result becomes quite important when the classifier is considered with the realistic parameters.  

Figs.~\ref{fig:Fig5} (b) and (d) exhibit that the classifier converges to the linear combination of the three reservoir states pointed out in the plot. In this specific example, the coupling strengths of the input channels are equal and the system qubit state reaches the unit fidelity where the target state is the linear combination of the given reservoir states with equal probabilities confirming that  Eqs. (\ref{Eq:convex}) and (\ref{Eq:ConvexL}) applies. 
More specifically, the steady state of the system qubit is $\varrho_m=p_1\ket{\psi_{\theta_1}}\bra{\psi_{\theta_1}}+p_2\ket{\psi_{\theta_2}}\bra{\psi_{\theta_2}}+p_3\ket{\psi_{\theta_3}}\bra{\psi_{\theta_3}}$ where $\sum_{i=1}^N p_i=1$. The probabilities experiencing from each channel are equal $p_1=p_2=p_3=1/3$ as the coupling of the system to the channels $J_1=J_2=J_3$ were set equal. This state specifies a steady magnetization, that is, the classifier returns a binary decision for the  three channel input case. Moreover, beyond the dynamical analysis of the classification process for two specific three input states, we also performed calculations for random input channel triple states parametrized by geometrical qubit angles $\theta$. The results are visualized as three dimensional parameter space $\theta_i$ where each triples of $\theta$ are generated randomly between $0$ and $\pi$ with mean $0$ and variance $1$. As clear in Fig.~\ref{fig:Fig5} (f) the data instances are linearly separable and the classifier operates properly also in the three input case. In general, for $N$ input channels the steady state of the classifier is $\varrho_m=\sum_{i=1}^N \ket{\psi_{\theta_i}}\bra{\psi_{\theta_i}}$ where steady state spin polarization $\langle \sigma_z \rangle_{ss}= \text{Tr}[\varrho_m\sigma_z]_{ss}$ always returns a binary decision.

\subsection{Physical model}

In this subsection, we propose a physical model  for the implementation of the quantum classifier. We choose the superconducting circuits~\cite{wendin_quantum_2017} as the physical model represents the theoretical example contains a single qubit in contact with two reservoir (ancilla) qubits. Again, the reservoirs are modelled by a repeated-interaction scheme and the physical qubits are the transmon qubits that interact through a resonator bus~\cite{koch_charge-insensitive_2007} in which also serves for qubit readout~\cite{bianchetti_dynamics_2009}.  

In general, the Hamiltonian of $N$ transmon qubits coupled via a coplanar waveguide (CPW) resonator reads 
\begin{align}\label{Transmon}
\mathcal{H}=&\omega_r\hat{a}^{\dagger}\hat{a}+\sum_{i=1}^{N} \left[ E_{c_i}(\hat{n}_i-n_{g_i})^2-E_{J_i}\cos \hat{\varphi}_i\right] \nonumber \\ 
& +\sum_{i=1}^{N} g_i \hat{n}_i (\hat{a}+\hat{a}^{\dagger}) 
\end{align}
where $\omega_r$ is the resonator frequency which is in essence, a quantum harmonic oscillator, $\hat{a}$ and $\hat{a}^{\dagger}$ are, respectively, the lowering and raising operators of the oscillator. Transmon qubit is a developed version of a charge qubit (Cooper pair box) based on the Josephson junction tunnelling device~\cite{makhlin_quantum-state_2001}. The second term in the Hamiltonian describes the charge qubits where $\hat{n}_i$ is the charge quanta number operator, $n_{g_i}$ is the offset charge and $\hat{\varphi}_i$ the quantized flux of qubit $i$. Here, $\varphi_i=\pi \Phi_i/\Phi_0$ where $\Phi_i$ is the tunable magnetic flux of each qubit and $\Phi_0$ is the elementary flux quanta. The capacitive energies $E_{c_i}$ and the Josephson energies $E_{J_i}$ of the qubits are set $E_{J_i}\gg E_{c_i} $ so that the qubits operate in the transmon regime in which they capacitively couple to the resonator by $g_i$. 
\begin{figure}
\includegraphics[width=3.0 in]{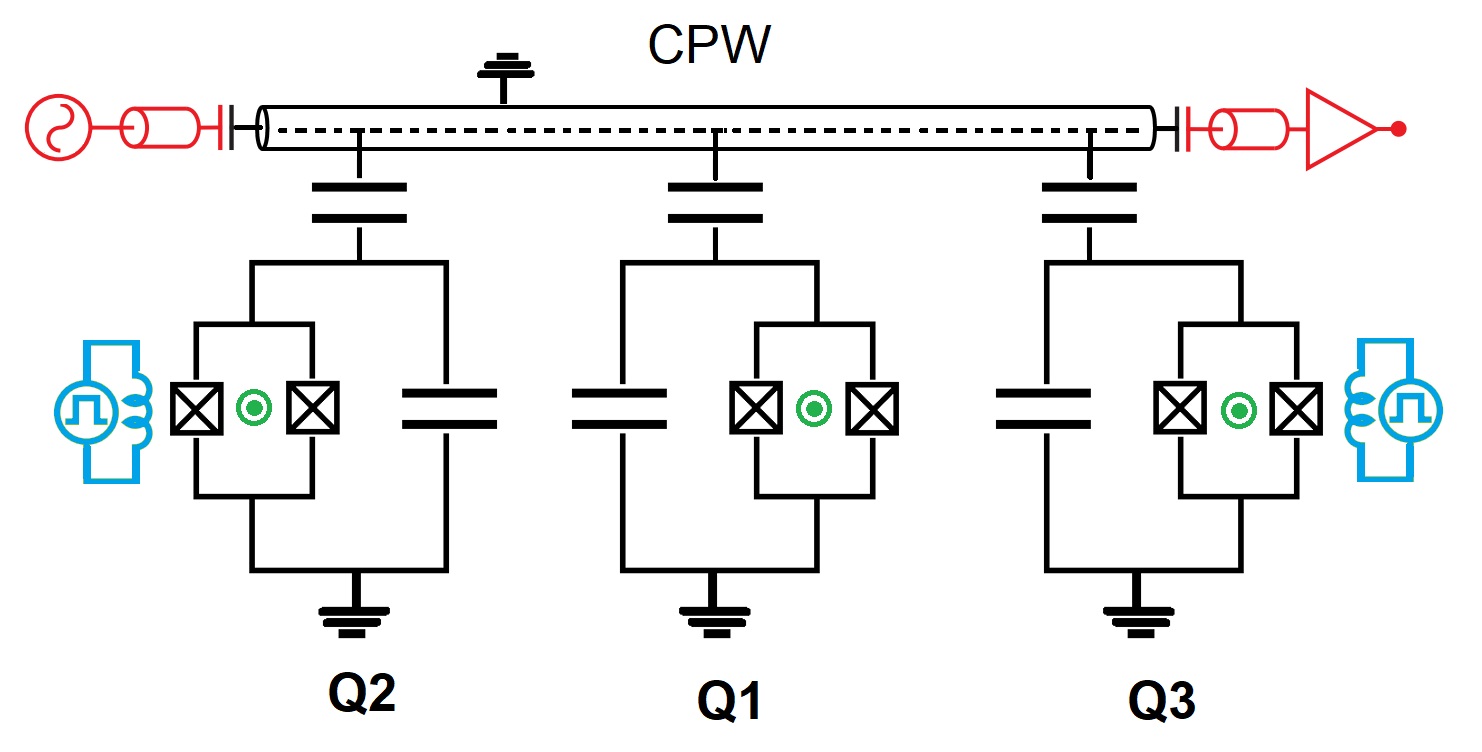}
\caption{ \label{fig:Fig6} (Colour online) Representation of the physical model of the quantum classifier by its Lumped-element circuit diagram. Three transmon qubits ($Q_1$ stands for the system qubit and $Q_2$ and $Q_3$ stand for the reservoir qubits) coupled to the superconducting CPW resonator that serves for both the readout of the qubits (red) and as a coupling bus. Green dots represent the flux tunability of each qubit that allow for the control of coupling to the bus via qubit frequencies. Control fields represented by Microwave lines (blue) acting on reservoir qubits $(Q_2, Q_3)$ are used for resetting and initialization of reservoir qubit states.} 
\end{figure} 
As clear in Eq.~(\ref{Transmon}) the desired interaction between qubits does not appear in the form as in Eq.~(\ref{Flip}). However, this type of interaction can directly be achieved by coupling the transmon qubits to the same resonator dispersively such as $\vert \Delta_{1,2,3} \vert=\vert \omega_{1,2,3}-\omega_r \vert\gg g_{1,2,3}$. In this scheme, the effective interaction between the qubits are, for instance $Q_1$ (the system qubit) and one of the ancilla qubits $Q_2$, described by ~\cite{majer_coupling_2007, filipp_multimode_2011}
\begin{equation}\label{Eq:couple}
J_{1,2}=\frac{g_1 g_2}{2} \left( \frac{1}{\Delta_1}+\frac{1}{\Delta_2}\right)
\end{equation}
in which the interaction is achieved via virtual exchange of cavity photons. Note that when $\vert \omega_1-\omega_2 \vert \gg J_{1,2}$ the interaction is effectively turned off, that is, the coupling strength can effectively be controlled by tunning the transmon qubit frequencies. 

There are some specific requirements to implement the proposed classifier by means of the physical system expressed in Fig. 5. First, the qubits that mimic the reservoirs $Q_2$ and $Q_3$, should interact with the system qubit $Q_1$ and should never interact each other. Second, a successive switch on/off mechanism should be achieved between the interacting qubits in accordance to suitable qubit state preparation and reset scenarios. The first requirement can be easily achieved by tuning the reservoir qubit frequencies $\vert \omega_2-\omega_3 \vert\gg J_{2,3}$ largely dispersive. Therefore, one obtains an effective Hamiltonian 
\begin{align}
\mathcal{H}=&\frac{\omega_i}{2}\sum_{i=1}^3\sigma_z^i+(\omega_r+\chi_i\sum_{i=1}^3 \sigma_z^i)\hat{a}^{\dagger}\hat{a}\nonumber \\&+J_{1,i}\sum_{i=2,3}(\sigma_1^+\sigma_i^- +H.c.)
\end{align}  
where $\sigma_z^i$ and $\sigma_i^{\mp}$ are the Pauli operators acting on the subspace representing the first two levels of the $i$th superconducting qubit. Here, $\chi_i$ are the qubit-dependent resonator frequency shift where there is no energy exchange between dispersively coupled qubit-resonator pairs. However, the second requirement should be evaluated by care, taking the realistic parameters into account. The physical implementation of the repeated interaction model, that is, the realization of the switch on/off mechanism between the reservoir qubits and the resonator can be performed by the externally tunable magnetic flux $\Phi_i$ as shown in the caption of Fig. 5.~\cite{liao_single-particle_2010}. The qubit-CPW coupling can be switched off by detuning the the qubit with the resonator very largely by using the flux bias and the coupling can be reproduced by again tuning $\Phi_i$ so that the desired dispersive coupling is achieved. 

In our scheme, the qubit-CPW coupling is switched on and off by repetitive steps. The time elapsed between two switching instants $t_i$ and $t_{i+1}$ is $T\leq t_{i+1}-t_i$ where $T=\tau_{int}+\tau_r+\tau_{pr}$. Here, we have several time scales where $\tau_r$ is the relaxation time of the qubit, $\tau_{int}$ is the qubit-CPW interaction time and $\tau_{pr}$ represents both, qubit reset and preparation times. In the scenario, the system transmon qubit $Q_1$ and the reservoir transmon qubits $Q_2$, $Q_3$  couple to a CPW resonator dispersively with strength $g$ that generates effective $J_{1,2}$ and $J_{1,3}$ couplings between $Q_1-Q_2$ and $Q_1-Q_3$ as discussed above. The reservoir transmon qubits are initially assumed to be prepared in their reservoir states before the switch-on interaction and $Q_1$ is prepared in any state such that $ \langle \sigma_z(0)\rangle=0$. At time $t_i=0$ the coupling between $Q_1$, $Q_2$,$Q_3$ and CPW is switched on. After $\tau_{int}$ the couplings are switched off and $Q_1$ and $Q_3$ are reset to their initial reservoir states after an elapsed $\tau_{pr}$ time. Hence, the system qubit is decoupled from the ancilla qubits and ready for the next time in the tensor product state. 
For reservoir qubits the relaxation time $\tau_r$ is much longer than the interaction time $\tau_r \gg \tau_{int}$  therefore has no effect on the reservoir qubits between any successive reset times. On the other hand, through a strong field, the qubit reset and preparation time $\tau_{pr}$ is much shorter than the interaction time $\tau_{pr}\ll \tau_{int}$~\cite{liao_single-particle_2010}. Then approximately, the relevant time scale between two successive switch-on operations is $t_{i+1}-t_i=T\simeq \tau_{int}$. 

Many  repetitions of the task described above, in principle can successfully simulate the proposed classifier model. However, there are some limitations on the achievement of the physical model. Possible preparation defects of the identical reservoir states should be taken into account. For a better performance analysis of the proposed physical model, one should encounter the experimental parameters to the calculations with the phenomenological decay rates. A comprehensive analysis of the CPW-qubits system can be carried out by a master equation approach with realistic parameters~\cite{kirchhoff_optimized_2018}. 

\begin{figure}
\includegraphics[width=3.2 in]{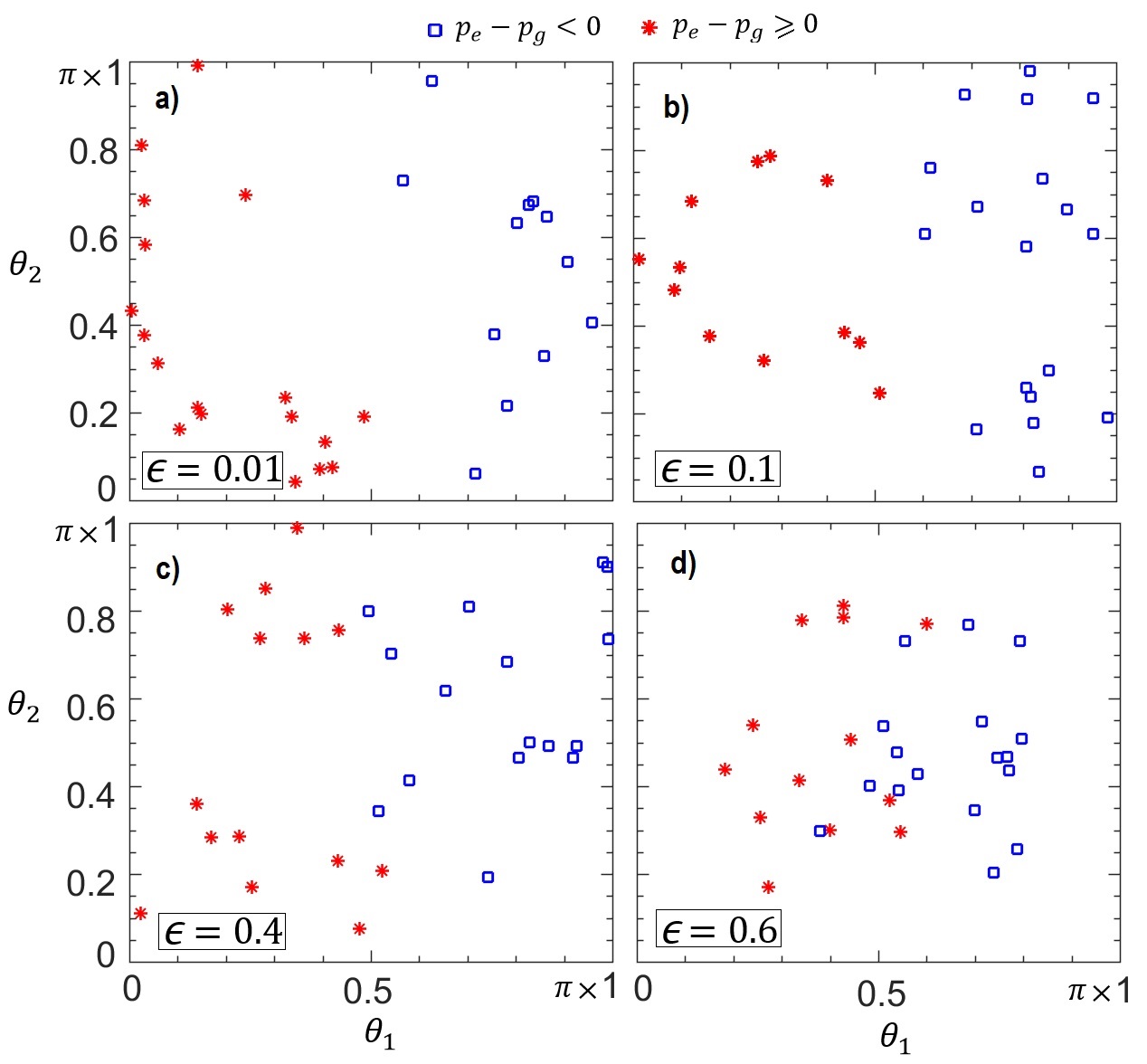}
\caption{ \label{Fig:fig7}  (Colour online) Deformation of the classified instances obtained by the physical model depending on the reservoir state preparation errors. Corresponding error rates are (a) $\epsilon=0.01$, (b) $\epsilon=0.1$, (c) $\epsilon=0.4$ and (d) $\epsilon=0.6$ with corresponding random rates $\eta=\epsilon/4$. Remained parameters are given in the text.} 
\end{figure} 

Here, we repeat the calculations of the proposed classifier with two reservoir qubits (expressed in Section 3.1) taking the reservoir qubit state preparation errors into account in order to see how the system is robust against errors. Typically, for superconductor circuit experiments in the weak coupling regime, the resonator frequency is $\omega_r\sim 1-10$ GHz and the qubit resonator coupling is $g\sim 1-500$ MHz~\cite{wendin_quantum_2017,majer_coupling_2007}. We choose the resonator frequency $\omega_r=8.625$ GHz and the $Q_1$, $Q_2$, $Q_3$ qubit frequencies as, respectively, $\omega_1/2\pi=6.2$ GHz, $\omega_2/2\pi=4.052$ GHz and $\omega_3/2\pi=7.518$ GHz where we obtain the effective couplings between the system qubit and the reservoir qubits as $J_{12}=J_{13}=48.9$ MHz by Eq.~(\ref{Eq:couple}). We also choose the reservoir-system interaction time $\tau_{int}=5$ ns. The reservoir qubit states were prepared as
\begin{equation}
\varrho_{\epsilon}=(1-\epsilon_{\eta})\ket{\psi_{\theta}}\bra{\psi_{\theta}}+\frac{\epsilon_{\eta}}{2}\mathbf{1}
\end{equation}    
before any interaction where $\mathbf{1}$ is the single qubit identity operator, $\ket{\psi_{\theta}}$ is the perfect reservoir state parametrized by Bloch angle $\theta$ and $\epsilon_{\eta}$ is the random error parameter. Here, $\epsilon_{\eta}=\epsilon\pm\eta$ and the random parameter $\eta$ models the errors of the identical reservoir state preparation. The reservoir states can be prepared by single qubit rotations. Current state-of-the-art allows for high fidelity logic gates by using single and two qubit rotations. For instance, two-qubit gates with $0.999-0.996$ fidelity corresponding to an infidelity $\epsilon=0.001-0.004$ were achieved for different types of multi-transmon qubits~\cite{sheldon_procedure_2016,deng_robustness_2017}.  

As shown in Fig.~\ref{Fig:fig7} (a)-(b) for relatively large error values $\epsilon=0.01$ or $\epsilon=0.1$ the physical machine can properly classify the data instances. However, for the exaggerated error values of $\epsilon$, (0.4 or 0.6 as in Fig.~\ref{Fig:fig7} (c)-(d)) the instances become inseparable. In these calculations we observe that the steady states were reached after $1500-2000$ collisions supporting the results in Fig.~\ref{Fig:fig2}. Note that, as the interaction time is $\tau_{int}=5$ ns for each collision, the required time to reach the steady state is $\sim 7.5-10\mu$s. As the spin polarization of the system qubit $Q_1$ is the recognizer of the proposed classification function of our study, energy relaxation time $T_1$ is relevant to the classifier performance. That is, $T_1$ of the system qubit should be larger than $\sim 10$ $\mu$s, the physical classifier response time for the proper functionality of the physical model. Recent studies report that energy relaxation time ranges between $T_1\sim 20-60$ $\mu$s depending on the coupling and the optimal noise suppressing pulse shape techniques of the transmon qubits ~\cite{kirchhoff_optimized_2018, deng_robustness_2017}. 

Beyond its feasibility, the physical model has advantageous features regarding its speed and the resource requirements. First, depending on  the current computer CPU capabilities, one can analyse that just like the typical CPU operations, the classical binary classifiers operate in the $m$s time range~\cite{bini_minimizing_2009}. As stressed above, the quantum classifier can process information in the $\mu$s range, that is, the proposed physical model is three orders of magnitude faster than its classical analogues. Second, note that the proposed classifier process quantum information and recently  proposed quantum classifiers rely on the circuit model of the universal quantum computing~\cite{schuld_simulating_2015,yamamoto_simulation_2018} in which requires  multi-qubit output registers for multi-qubit inputs. However, the proposed quantum perceptron model achieves the binary classification task only by a single qubit output regardless of the number of input channels. Therefore, the proposed quantum classifier in which has the speed superiority in comparison to the classical classifiers, is advantageous also in terms of using the resources in comparison to the other quantum  classifiers. 

Finally, we would like to mention the possibility of the physical model to implement on the IBM quantum computer. Currently IBM builds a universal quantum computer using a superconductor circuit architecture composed of transmon qubits through IBM Quantum  Experience project (IBMQX)~\cite{research_ibm}.  Just like proposed classifier, IBMQX architecture depends on the microwave transmon qubits coupled via CPW resonators. However, the proposed repeated-interaction scheme with frequency tunable transmon qubits, can not be directly applied to IBM architecture as they couple the qubits by exploiting cross-resonance effect in which the qubit frequencies are fixed ~\cite{chow_simple_2011}. But note that universal quantum computers, in principle, can simulate open quantum dynamics through repeated algorithms~\cite{wang_quantum_2011, kretschmer_collision_2016}. This type of universal quantum simulation which is the basic motivation of the  quantum computers, is called digital quantum simulation~\cite{zalka_christof_simulating_1998, heras_digital_2014, salathe_digital_2015}. The limitation of this scheme is that the quantum digital simulation is an approximation of the actual system Hamiltonian and the cost to pay the approximation is the Hamiltonian decomposition errors~\cite{trotter_product_1959, knee_optimal_2015}.

\section{Conclusion}
We propose a general and simple open quantum model for quantum data classification. Repeated interactions were chosen to model open system dynamics for faithful representation of arbitrary states of information reservoirs. It's numerically demonstrated that a single qubit quantum system weakly coupled to quantum reservoirs with quantum information is capable of classifying the input data. Two limit cases with fixed reservoir states and fixed couplings were examined in order to clearly demonstrate the equilibrium state response of the system. 

We show that the steady state response of the system has a non-linear activation function-like behaviour for both linear variation of coupling and reservoir state parameters. Three input channel example was also demonstrated. A possible application of the proposed model by superconducting circuits was discussed in detail and the physical performance of the classifier was examined regarding the defects. Considering the experimental parameters, we conjectured that the physical model operates three orders of magnitude faster than the classical counterparts. Thus, we have shown that an open quantum system, which is generally considered to have no useful information in its equilibrated state, is a natural quantum data classifier.

\section{Acknowledgements}
Authors acknowledge support from Istanbul Technical University. This research did not receive any specific grant from funding agencies in the public, commercial, or not-for-profit sectors.
 

\balance



\end{document}